\documentclass[iop]{emulateapj}
\usepackage{color}
\usepackage{natbib}
\shorttitle{CARMA CO(\textit{J} = $2-1$) Observations of Betelgeuse's CSE}
\shortauthors{O'Gorman et al.}

\begin{document}

\title{CARMA CO(\textit{J} = 2 -- 1) OBSERVATIONS OF THE CIRCUMSTELLAR ENVELOPE OF BETELGEUSE}

\author{Eamon O'Gorman\altaffilmark{1}, Graham M. Harper\altaffilmark{1}, Joanna M. Brown\altaffilmark{2}, Alexander Brown\altaffilmark{3}, Seth Redfield\altaffilmark{4},\\ Matthew J. Richter\altaffilmark{5}, and Miguel A. Requena-Torres\altaffilmark{6}}
\altaffiltext{1}{School of Physics, Trinity College Dublin, Dublin 2, Ireland.}
\altaffiltext{2}{Harvard-Smithsonian Center for Astrophysics, 60 Garden Street, MS-78, Cambridge, MA 02138, USA.}
\altaffiltext{3}{Center for Astrophysics and Space Astronomy, University of Colorado, Boulder, CO 80309-0389, USA}
\altaffiltext{4}{Astronomy Department, Van Vleck Observatory, Wesleyan University, Middletown, CT 06459, USA}
\altaffiltext{5}{Physics Department, UC Davis, 1 Shields Avenue, Davis, CA 95616, USA}
\altaffiltext{6}{Max-Planck-Institut f\"ur Radioastronomie, Auf dem H\"ugel 69, 53121 Bonn, Germany}


\begin{abstract}
We report radio interferometric observations of the $\rm{{}^{12}}$C$\rm{{}^{16}}O$ 1.3\,mm \textit{J} = $2-1$ emission line in the circumstellar envelope of the M supergiant $\alpha$ Ori and have detected and separated both the S1 and S2 flow components for the first time. Observations were made with the Combined Array for Research in Millimeter-wave Astronomy (CARMA) interferometer in the C, D, and E antenna configurations. We obtain good $u-v$ coverage (5--280\,k$\lambda$) by combining data from all three configurations allowing us to trace spatial scales as small as $0\arcsec.9$ over a $32\arcsec$ field of view. The high spectral and spatial resolution C configuration line profile shows that the inner S1 flow has slightly asymmetric outflow velocities ranging from $-9.0\>{\rm km\>s}^{-1}$ to $+10.6\>{\rm km\>s}^{-1}$ with respect to the stellar rest frame. We find little evidence  for the outer S2 flow in this configuration because the majority of this emission has been spatially-filtered (resolved out) by the array. We also report a SOFIA-GREAT CO(\textit{J} = $12-11$) emission line profile which we associate with this inner higher excitation S1 flow. The outer S2 flow appears in the D and E configuration maps and its outflow velocity is found to be in good agreement with high resolution optical spectroscopy of K\,I obtained at the McDonald Observatory. We image both S1 and S2 in the multi-configuration maps and see a gradual change in the angular size of the emission in the high absolute velocity maps. We assign an outer radius of 4$\arcsec$ to S1 and propose that S2 extends beyond CARMA's field of view (32$\arcsec$ at 1.3 mm) out to a radius of 17$\arcsec$ which is larger than recent single-dish observations have indicated. When azimuthally averaged, the intensity fall-off for both flows is found to be proportional to \textit{R}$^{-1}$, where \textit{R} is the projected radius, indicating optically thin winds with $\rm{\rho \propto \textit{R}^{-2}}$. 
\end{abstract}


\section{INTRODUCTION}
The circumstellar envelope (CSE) of the M2 Iab supergiant Betelgeuse ($\alpha$ Orionis) is a proving ground for ideas and theories of mass-loss from oxygen-rich M supergiants. Currently it is losing mass at a respectable rate $\sim 3\times 10^{-6} \ M{}_{\odot}$ yr${}^{-1}$ \citep{1986ApJ...306..605G, 1994ApJ...424L.127H,harper_2001}, as it has been over the past $\sim$ 1000 yr. Most of the optically thin silicate dust lies beyond $\sim 46$ stellar radii \citep{1994AJ....107.1469D} and dust is, therefore, unlikely to be responsible for the bulk mass-loss. This raises the important point that if the mass-loss from Betelgeuse is not a result of dust then perhaps the same mechanisms that are responsible might also be active in the more dusty later M-type supergiants. 

\begin{deluxetable*}{cccccc}
\tabletypesize{\scriptsize}
\tablewidth{0pt} 
\tablecaption{CARMA Observations}
\tablehead{\colhead{Observation}		                                &
                 \colhead{Configuration}				&
          	\colhead{Time on Source}	      			&
	\colhead{Flux}			    	     	&
           	\colhead{Phase}                                       		&
	\colhead{Image Cube\tablenotemark{a}}		\\
	\colhead{Date}					&
           	\colhead{}                				& 
	\colhead{(hr)}                                          		& 
	\colhead{Calibrator}                                          		& 
	\colhead{Calibrators}                                          		&
	\colhead{Dynamic Range\tablenotemark{b}}		}
\startdata
2007 Jun 18 	& D & 0.9 & 0530+135	& 0530+135, 0532+075 	&  22.8 \\
2007 Jun 21 	& D & 3.0 & 0530+135	& 0530+135, 0532+075 	&  22.7 \\
2007 Jun 24 	& D & 2.1 & 0530+135	& 0530+135, 0532+075 	&  26.1 \\
2007 Jun 25 	& D & 2.4 & 0530+135	& 0530+135, 0532+075 	&  30.2 \\
2009 Jul 7	& E & 3.2 & 3C120 		& 3C120, 0532+075	& 30.1 \\
2009 Nov 5	& C & 1.2 & 3C120 		& 3C120, 0532+075 	& 17.3 \\
2009 Nov 09 	& C & 3.0 & 3C120 		& 3C120, 0532+075 	& 27.2 \\
2009 Nov 15	& C & 1.0 & 3C120 		& 3C120, 0532+075 	& 17.8 \\
2009 Nov 16	& C & 3.2 & 3C120 		& 3C120, 0532+075 	& 32.0  \\
All		& C & 8.4	& \nodata 	& \nodata 		& 43.8 \\
All 		& D & 8.4 & \nodata 	& \nodata 		& 31.9 \\
All 		& Multi-configuration & 20.0 & \nodata & \nodata 	& 52.3 
\enddata
\tablenotetext{a}{Low spectral resolution (i.e., channel width of $1.3\>{\rm km\>s}^{-1}$).}
\tablenotetext{b}{The peak emission of the image cube divided by the root mean square of the residual image.}
\label{tab:tab1}
\end{deluxetable*}
\cite{1984ApJ...284..238H} constructed a WKB Alfv\'{e}n wave-driven model for Betelgeuse's chromosphere and wind that reproduced reasonably well the observed chromospheric emission fluxes and mass-loss rate. However, multi-wavelength centimeter-continuum radio observations made with the Very Large Array (VLA) by \cite{1998Natur.392..575L} revealed that the atmosphere was much cooler than the Alfv\'{e}n wave-driven model predicted \citep{harper_2001}. Although  new nonlinear Alfv\'{e}n wave models have been computed by \cite{2000ApJ...528..965A}, there are currently no theoretical models that make specific predictions for {\em both} the dynamic and thermodynamic state of the mass outflow. Radiation pressure on atoms and molecules is another potential contributing candidate as a mass-loss mechanism and so spatial and dynamical studies of molecules are a fruitful line of investigation, especially in relation to eventual formation of dust. Such studies also allow us to calculate the time scales on which certain mass-loss episodes have occurred, and these can then be compared to the time scales of potential mass-loss initiators such as convection or magnetic dynamo cycles.

The study of CO molecules in the CSE of Betelgeuse began with the detection of 4.6\,$\mu$m ro-vibrational absorption lines of $\rm{{}^{12}C^{16}O}$ and $\rm{{}^{13}C^{16}O}$ by \cite{1979ApJ...233L.135B} who identified two absorption features, implying two distinct structures within the overall outflow. One component, known as S1, has a Doppler shift of $9\>{\rm km\>s}^{-1}$ toward us with \textit{T}$_{\rm{exc}}\,\simeq$\,200\,$\rm{K}$, $v_{\rm{turb}}\,\simeq\,4$\,km\,s${}^{-1}$ and \textit{N}$_{\rm{^{12}C^{16}O}}=4.7\times 10^{17}\>{\rm cm}^{-2}$. The second faster component, known as S2, has a Doppler shift of $16\>{\rm km\>s}^{-1}$ towards us with \textit{T}$_{\rm{exc}} \simeq$ 70 $\rm{K}$, $v_{\rm{turb}}\simeq 1$\,km\,s${}^{-1}$ and \textit{N}$_{\rm{^{12}C^{16}O}}=1.2\times 10^{16}\>{\rm cm}^{-2}$. The S1 feature with its higher column density was well known from atomic absorption line studies \citep[e.g.,][]{1962ApJ...136..844W} and both features had been detected in high spectral resolution atomic Na and K absorption profiles \citep{1975ApJ...199..427G}. $\rm{{}^{12}C^{16}O}$ was subsequently detected at 230\,GHz in the $J=$ $2-1$ rotational emission line by \cite{1980ApJ...242L..25K}, although a search for SiO($J=$ $2-1$) by \cite{1978ApJ...221..854L} had been unsuccessful. The weaker $\rm{{}^{12}C^{16}O}$($J=$ $1-0$) line was tentatively detected by \cite{1985ApJ...292..640K} with a 7\,m dish which had a half-power beamwidth (HPBW) of 100$\arcsec$.

\cite{1987ApJ...313..400H} presented a higher signal to noise $\rm{{}^{12}C^{16}O}$($J=$ $2-1$) observation of Betelgeuse's CSE with an HPBW of 32$\arcsec$ and found some evidence for an S2 radius of about $16\arcsec$ by comparing the  ($2-1$)/($1-0$) intensities. However,  a 30\,m Institut de Radioastronomie Millim\'etrique (IRAM) $J=$ $2-1$ line profile was later presented by  \cite{1994ApJ...424L.127H} and looked remarkably similar, even though it was observed with a smaller 12$\arcsec$ HPBW. The profile did not show the horned wing signature expected if it had been resolved apparently in conflict with the previous S2 radius estimate.

Here we present the results of an interferometric study of the rotational $\rm{{}^{12}C^{16}O}$($J=$ $2-1$) emission line made using three Combined Array for Research in Millimeter-wave Astronomy (CARMA) configurations with HPBWs of 0$\arcsec$.9 (C), 2$\arcsec$.1 (D), and 4$\arcsec$.4 (E) designed to explore the S1 and S2 flows at these spatial scales. Preliminary results of the D configuration observations have been presented in \cite{2009AIPC.1094..868H} as part of a multi-wavelength study of CO surrounding $\alpha$ Ori. We also present a supporting SOFIA CO(\textit{J} = $12-11$) line profile in addition to high spectral resolution observations of the K\,I\,7699\,\AA \ line. In Section 2 the observations and data reduction techniques are discussed and in Section 3 the results of the spectra and image maps are presented. Discussions and conclusions are presented in Section 4 and Section 5, respectively.

\section{OBSERVATIONS AND DATA REDUCTION}

The millimeter observations were made with the 15 element CARMA interferometer \citep{2004ASPC..314..768S} which is located at Cedar Flat in eastern California at an elevation of 2200\,m. The array consists of nine 6.1\,m antennas and six 10.4\,m antennas formerly from the Berkeley Illinois Maryland Association (BIMA) and the Owens Valley Radio Observatory (OVRO) arrays respectively. Table \ref{tab:tab1} summarizes the various observations which span the period 2007 May$-$2009 November. The observations were carried out in the C, D, and E configurations and consist of on-source profiles of the $\rm{{}^{12}}$C$\rm{{}^{16}}$O($J=$ $2-1$) line which has a rest frequency of 230.538\,GHz (1.3\,mm). The baseline length spans over 26$-$370\,m (C array), 11$-$148\,m (D array), and 8.5$-$66\,m (E array) providing HPBWs of 0$\arcsec$.9, 2$\arcsec$.1, and 4$\arcsec$.4 respectively at 1.3\,mm. The HPBW of the individual 10.4\,m antennas is $\sim$ 32$\arcsec$ at the observed frequency.

The CARMA correlator takes measurements in three separate bands, each having an upper and lower sideband. One band was set to the low-resolution 468\,MHz bandwidth mode (15 channels of 31.25\,MHz each) to observe continuum emission and was centered on the line. The other two bands were configured with 62\,MHz and 31\,MHz bandwidth across 63 channels (with a resolution of $1.3\>{\rm km\>s}^{-1}$ and $0.65\>{\rm km\>s}^{-1}$, respectively) and were also centered on the line. The line was measured in the upper sideband in the C and E array and in the lower sideband in the D array.

Bandpass and phase calibration were performed using 3C120 and 0530+135. 0532+075 was used as a secondary phase calibrator to determine the quality of the phase transfer from the primary phase calibrator. The observing sequence was to integrate on the primary phase calibrator for $\sim$ 2.5 minutes, the target for $\sim$ 18 minutes, and the secondary phase calibrator for $\sim$ 2.5 minutes. The cycle was repeated for each track which lasted between 1.5 hr and 5 hr. Absolute flux calibration was carried out with 0530+135 and 3C120 using the continuously updated CARMA flux catalog to obtain their flux values at each observation.

The raw data were smoothed by a Hanning filter within MIRIAD\footnote{Multichannel Image Reconstruction, Image Analysis and Display, \url{http://www.atnf.csiro.au/computing/software/miriad/}} and then exported into FITS format so that it could be analyzed with the CASA\footnote{Common Astronomy Software Applications, \url{http://casa.nrao.edu/}} data reduction package. All calibration and imaging was carried out within CASA. The image cubes  were multi-scale  CLEANed down to the 3$\rm{\sigma}$ threshold using natural weighting and were corrected for primary beam attenuation. The \textit{multiscale} algorithm \citep{2008AJ....136.2897R} within CASA was set to four unique scales; the largest corresponding to the largest structures visible in individual channel maps. Each scale was approximately set to three times smaller than the preceding scale. 

Each of the three CARMA configurations sample a different range of spatial frequencies; the range of which is dependent upon the maximum and minimum baselines ($b_{\rm{max}}$ and $b_{\rm{min}}$) of each configuration. The sources we are observing are extended and therefore it is necessary to consider the response of each CARMA configuration to this extended emission. For any array configuration, emission with angular scales of $\sim \lambda/b_{\rm{min}}$ or greater is not reproduced in the maps \citep{1999ASPC..180.....T} and this scale is often used as a guide for the \textit{resolving out scale} or \textit{maximum scale} of an array configuration. To obtain a more robust estimate of the largest angular scale that can be accurately imaged in the high spatial resolution C configuration maps we computed the visibilities of an extended emission feature (whose spatial extent was set to that of the primary beam) using CASAs simulation tool, \textit{simdata}. This tool then produced a CLEANed image of these visibilities from which we calculated the resolving out scale to be $\sim$ 6$\arcsec$ (i.e., $0.6\lambda/b_{\rm{min}}$). Ultimately, combining the data from these three configurations allows the missing short spacings from the extended C configuration to be recovered while maintaining its high spatial resolution.

\section{RESULTS} 

Betelgeuse is a  semi-regular variable and its radial velocity exhibits variability on time scales ranging from short 1.5 year periods as suggested by \cite{1931PWasO..15..178S} to longer 5.8 year periods (Jones, 1928). Its radial velocity amplitudes are also known to vary by at least $\pm 3\>{\rm km\>s}^{-1}$ \citep{1989AJ.....98.2233S} making it difficult to determine a precise value for the stellar center-of-mass radial velocity. In this study we adopt a heliocentric radial velocity of $+20.7\>{\rm km\>s}^{-1}$ ($v_{\rm{lsr}}$ = $4.8\>{\rm km\>s}^{-1}$); a value adopted by \citet{2008AJ....135.1430H} and is based on the mean values of Jones (1928) and \cite{1933CMWCI.464....1S}. All spectra are plotted with respect to the stellar center-of-mass rest frame.

\subsection{CO($J=$ $2-1$) Spectra} \label{results1} 

\begin{figure}
\includegraphics[trim=90pt 60pt 45pt 50pt, clip, width=8.0cm, height=14.0cm]{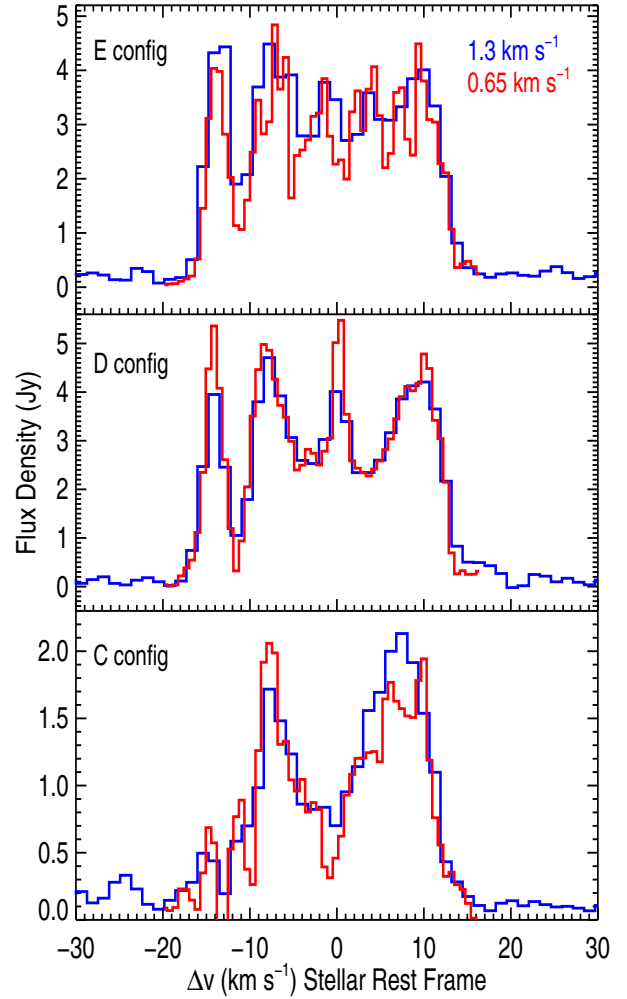}
\caption{Spectra integrated over a radius of 5$\arcsec$ for each array configuration image cube. The blueshifted emission component between $-16.0\>{\rm km\>s}^{-1}$ and $-10.0\>{\rm km\>s}^{-1}$ is almost resolved out in the C configuration image cube spectrum. The red and blue lines correspond to the high and low spectral resolution data respectively.\label{fig1}}
\label{fig:fig1}
\end{figure}

The spectrum for each individual configuration image cube (which are composed of all the appropriate configuration tracks listed in Table \ref{tab:tab1}) along with the multi-configuration image cube can be used to obtain information on the kinematics of the S1 and S2 flows. The spectra corresponding  to the C, D, and E configuration image cubes are plotted in Figure \ref{fig:fig1} for both the high ($0.65\>{\rm km\>s}^{-1}\> \rm{bin}^{-1}$) and low ($1.3\>{\rm km\>s}^{-1}\> \rm{bin}^{-1}$) spectral resolution data and were obtained by integrating all emission within a circular area of radius 5$\arcsec$ centered on the source. The high and low spectral resolution modes allow two independent sets of spectra to be measured for each observation and thus provide a good check on the data quality. The high resolution  spectra (channel width = $0.65\>{\rm km\>s}^{-1}$) give the best measure of S1 and S2 kinematics and therefore all outflow velocities are derived from these spectra.

The E configuration image cube spectrum has a total line width of $29.2\>{\rm km\>s}^{-1}$ and the low spectral resolution profile contains a steep blue wing emission feature between $-16.0\>{\rm km\>s}^{-1}$ and $-11\>{\rm km\>s}^{-1}$ and a more flat-topped feature between $-10.3\>{\rm km\>s}^{-1}$ and $-13.2\>{\rm km\>s}^{-1}$. This steep emission wing shows that the turbulence in the flow is less than or equal to the velocity bin size. The blue wing in the high resolution profile matches the lower resolution profile well but the  remainder of the profile looks more complex than the flat-topped feature seen in the lower resolution profile. The profile shape of the CO($J=$ $2-1$) line has been well documented by previous single-dish observations \citep[e.g.,][]{1980ApJ...242L..25K, 1987ApJ...313..400H} and, out of our three individual configuration spectra, we expect the most compact E configuration spectra to resemble these single-dish measurements the closest due to its better sampling of the inner $u-v$ plane and consequent sensitivity to extended structures. This indeed turns out to be the case when we compare our three individual configuration spectra to those previous single-dish profiles. The blue wing emission feature appears again in the D configuration spectrum at the same velocities as those in the E configuration spectrum but the remainder of the profile appears quite different. Between $-10.3\>{\rm km\>s}^{-1}$ and $+13.2\>{\rm km\>s}^{-1}$ the D configuration spectrum is dominated by a blue wing at $\sim$ $-10.0\>{\rm km\>s}^{-1}$, a red wing at $\sim$ $+13.0\>{\rm km\>s}^{-1}$ and an emission feature at $\sim$ $0\>{\rm km\>s}^{-1}$. 

The line profile has a much lower flux in the high spatial resolution C configuration spectrum due to its lack of sensitivity to extended structure. The blueshifted emission feature located between $-16.0\>{\rm km\>s}^{-1}$ and $-11.0\>{\rm km\>s}^{-1}$ in the E and D configuration spectra is almost completely spatially filtered by the extended C configuration. This component of the line has previously been associated with the outer S2 flow \citep{1987ApJ...313..400H} and as the majority of it has been spatially filtered by our C configuration we expect even less contribution from the S2 flow at lower absolute velocities still. For the redshifted line emission we again expect the majority of the S2 contribution to be spatially filtered, so we conclude that the majority of the emission in the C configuration spectrum emanates from the inner S1 flow. The spectrum is double peaked with the blue and redshifted wings extending to $-9.0\>{\rm km\>s}^{-1}$ and $+10.6\>{\rm km\>s}^{-1}$ respectively, and we define these as the outflow velocities of S1. As discussed in Section 2, the C configuration has a resolving out scale of $\sim$ 6$\arcsec$ at 1.3 mm and so is not sensitive to angular scales larger than this. If the emission between $-9.0\>{\rm km\>s}^{-1}$ and $+10.6\>{\rm km\>s}^{-1}$ in the C configuration spectrum appeared as a flat-topped profile then we could conclude that the S1 flow lies within a radius of 3$\arcsec$ from the star. Clearly however, the lower absolute velocity components of this profile have been spatially filtered so we conclude that the radial extent of the S1 from the star is greater than 3$\arcsec$. If we assume that the S1 flow would produce a top-hat line profile were it not for the resolving out effects of the interferometer, then its integrated line flux is 3.1 $\times$ 10${}^{-19}$\,W\,m${}^{-2}$.

\begin{figure}
\includegraphics[scale=0.75, angle=90, width=13.0cm, height=11cm, trim=20pt 50pt 20pt 40pt]{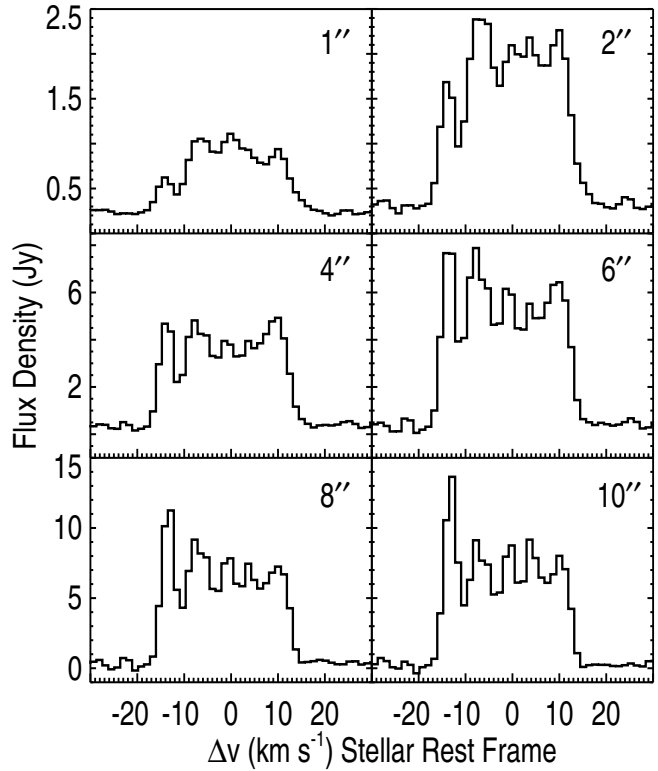}
\caption{Spectral profiles of the low spectral resolution multi-configuration image cube for circular extraction areas of radius 1$\arcsec$, 2$\arcsec$, 4$\arcsec$, 6$\arcsec$, 8$\arcsec$, and 10$\arcsec$. The signal to noise of the line profile reduces at larger extraction areas as more noise is included from the outer regions of the channel maps.}
\label{fig:fig2}
\end{figure}

To obtain the most robust value for the S2 outflow velocities we examine the high spectral resolution multi-configuration image cube spectrum which is composed of all tracks from all three configurations. It is worth stressing that by analyzing the multi-configuration image cube we make the assumption that the physical properties of all three components (i.e., $\alpha$ Ori, S1, and S2) have not changed over the total observation period (i.e., $\sim$ 2.5 yr). The profile is found to have a total line width of 28.6 $\pm$ $0.7\>{\rm km\>s}^{-1}$, which is in close agreement with previous single-dish observations of the line where values of 30.6 $\pm$ $2.5\>{\rm km\>s}^{-1}$ and $28.6\>{\rm km\>s}^{-1}$ were reported by \cite{1980ApJ...242L..25K} and \cite{1987ApJ...313..400H}, respectively. The centroid velocity of the spectrum is  -1.1 $\pm$ $0.7\>{\rm km\>s}^{-1}$ ($v_{\rm{lsr}}$ = 3.7 $\pm$ $0.7\>{\rm km\>s}^{-1}$) which is again in close agreement with \cite{1980ApJ...242L..25K} and \cite{1987ApJ...313..400H} values of $v_{\rm{lsr}}$ = 3.0 $\pm$ $2.5\>{\rm km\>s}^{-1}$ and $v_{\rm{lsr}}$ = 3.7 $\pm$ $0.4\>{\rm km\>s}^{-1}$ respectively. The integrated line flux is 1.5 $\times 10^{-18}$\,W\,m$^{-2}$ of which approximately 20\% emanates from the S1 flow.

The outflow velocities of S2 are $-15.4\>{\rm km\>s}^{-1}$ and $+13.2\>{\rm km\>s}^{-1}$ which, like the S1 flow, are slightly asymmetric but in the opposite sense. Note that the S1 and S2 outflow velocities are dependent on the adopted radial velocity of Betelgeuse. If, for instance, we instead adopt a radial velocity of 21.9 km s${}^{-1}$ \citep{2005A&A...430..165F} then the S2 outflow velocities become even more asymmetric ($-16.6$ and $+12.0\>{\rm km\>s}^{-1}$) while the S1 outflow becomes less so ($-10.2$ and $+9.4\>{\rm km\>s}^{-1}$). Both S1 and S2 therefore cannot have spherically symmetric outflow velocities regardless of the adopted stellar radial velocity. Adopting a mass of 18\,$M_{\odot}$ and a radius of 950\,$R_{\odot}$ \citep{2008AJ....135.1430H} then the escape velocity for Betelgeuse is $85\>{\rm km\>s}^{-1}$ which is much greater than the S1 and S2 outflow velocities. This indicates that the majority of the stellar mass-loss mechanism's energy goes into lifting the CO molecules out of the gravitational potential and not into their outflow velocities. These outflow velocities are greater than the adiabatic hydrogen sound speed, which, if we assume that the gas temperature is the same as the excitation temperature, are $1.7\>{\rm km\>s}^{-1}$ and $1\>{\rm km\>s}^{-1}$ for S1 and S2, respectively. 

The spectra in Figure \ref{fig:fig2} are taken from the low-resolution multi-configuration image cube using circular extraction areas ranging in radius from 1$\arcsec$ to 10$\arcsec$ and demonstrates how the line profile changes over these different extraction areas. The most striking change in the line profile is the change in appearance of the extreme blue wing. At small extraction radii where we sample the most compact emission, the feature is weak in comparison to the rest of the line but becomes more dominant as we begin to sample more of the extended emission. This indicates that even the high absolute velocity components of the S2 flow have extended emission and this is why they are almost completely spatially filtered by CARMA's C configuration. The large reduction of flux at $-11\>{\rm km\>s}^{-1}$ suggests that there is more material moving toward the observer than at other lower absolute velocities indicating a non-isotropic (or non-spherical) S2 flow. This suggests a more sheet-like (flatter) structure rather than a spherical cap.

\subsection{Multi-configuration Image Cube} \label{results2} 

\begin{figure}[hbt!]
\centering
\mbox{
          \includegraphics[]{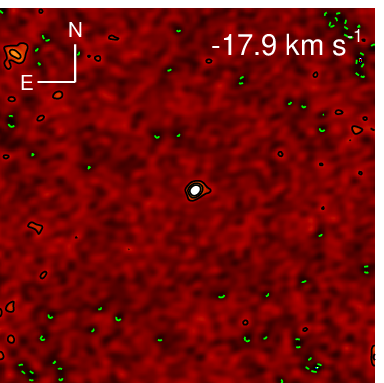}
          \includegraphics[]{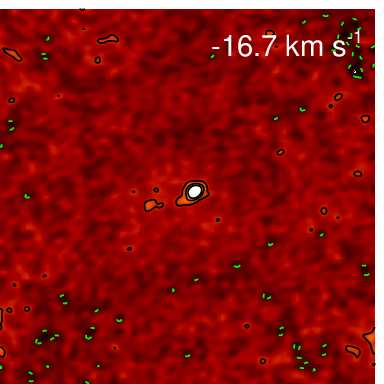}
          }
\\
\mbox{
          \includegraphics[]{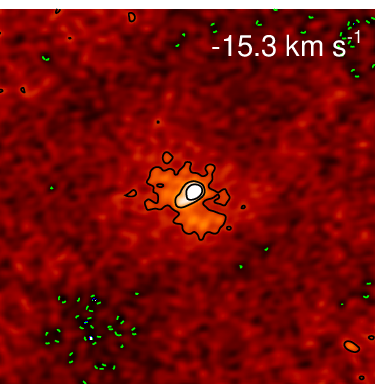}
          \includegraphics[]{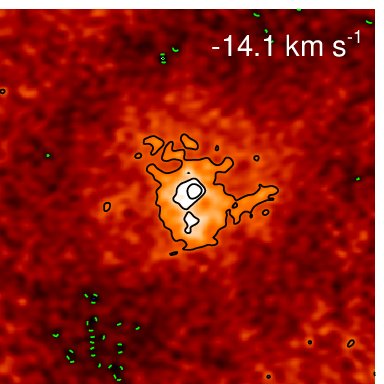}
          }
\\
\mbox{
          \includegraphics[]{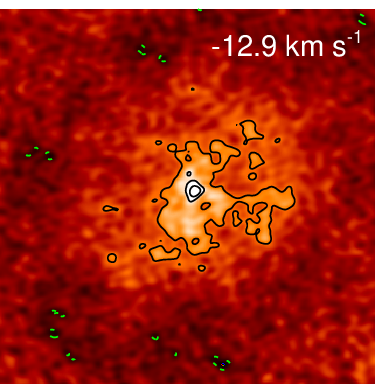}
          \includegraphics[]{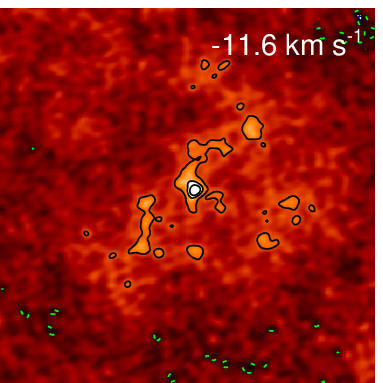}
         }
\\
\mbox{
          \includegraphics[]{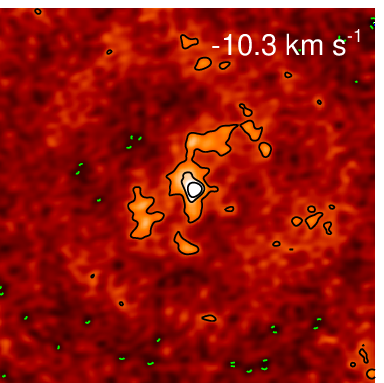}
          \includegraphics[]{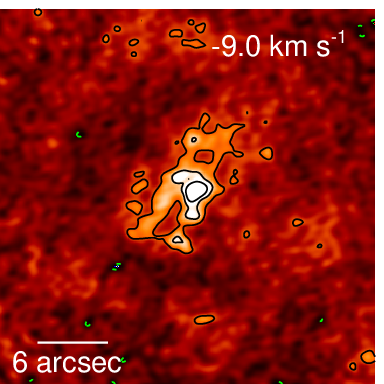}
         }
\includegraphics[trim=0pt 20pt 0pt 5pt]{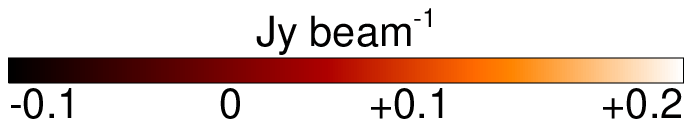}
\caption{Eight channel maps from the multi-configuration image cube ($\Delta v$ = $1.3\>{\rm km\>s}^{-1}$). The peak emission has been cut at 0.2 Jy beam${{}^{-1}}$ to emphasize the fainter emission. The color scale is linear and has been normalized to this maximum cutoff and minimum value of each channel. The contour levels are at $-2\sigma$, 2$\sigma$, 4$\sigma$, and 6$\sigma$ (1$\sigma$ $\sim$ 0.03 Jy beam$^{-1}$ but varies per channel). Dashed green lines represent negative contours.}
\label{fig:fig3}
\end{figure}

A subset of the blueshifted velocity channel maps of the low spectral resolution multi-configuration image cube is presented in Figure \ref{fig:fig3}. The first channel map at $-17.9\>{\rm km\>s}^{-1}$ shows just the compact unresolved continuum emission with no extended emission present. Between $-16.7\>{\rm km\>s}^{-1}$ and $-9.0\>{\rm km\>s}^{-1}$, we see evidence for the development of a classical shell signature for the S2 flow. We first sample the highest velocity components where the emission is relatively compact (i.e., between $-16.7\>{\rm km\>s}^{-1}$ and $-12.9\>{\rm km\>s}^{-1}$) and then sample lower radial velocity components where S2 becomes a faint ring (i.e., between $-11.6\>{\rm km\>s}^{-1}$ and $-9.0\>{\rm km\>s}^{-1}$). At lower velocities again, these rings disappear into the noise of the maps and possibly extend out beyond the primary beam at zero velocity when the rings should have maximum spatial extent. The emission from the channel maps between $-15.3\>{\rm km\>s}^{-1}$ and $-11.6\>{\rm km\>s}^{-1}$ corresponds to all the emission in the extreme blue wing component of the multi-configuration image cube line profile discussed in Section 3.1. We can see in Figure \ref{fig:fig3} that all of this emission is greater than the C configuration resolving out scale therefore confirming that our C configuration line profile is mainly composed of S1 emission. The shell formation signature of S2 is also apparent in the redshifted velocity channel maps between $+7.5\>{\rm km\>s}^{-1}$ and $+13.8\>{\rm km\>s}^{-1}$ but the emission appears weaker and the rings fainter therefore indicating that S2 is somewhat fragmented. 

The multi-configuration maps also show the central compact emission from the S1 flow at velocities between $-10.3\>{\rm km\>s}^{-1}$ and $+11.3\>{\rm km\>s}^{-1}$. This S1 emission can be seen in the final two maps of Figure \ref{fig:fig3} as a central slightly elongated emission feature surrounded by the fainter rings of the S2 flow. In the maps where both S1 and S2 are present the emission from S1 appears brighter than the emission from S2. The spatial extent of the S1 flow varies from channel map to channel map but appears to be larger than the 2$\arcsec$ value given by \cite{2009AJ....137.3558S}, who observed off-star wind scattered ro-vibration CO lines. 

\begin{figure}
\epsscale{1.0}
\includegraphics[trim=25pt 0pt 50pt 0pt, width=8.0cm, height=6.8cm]{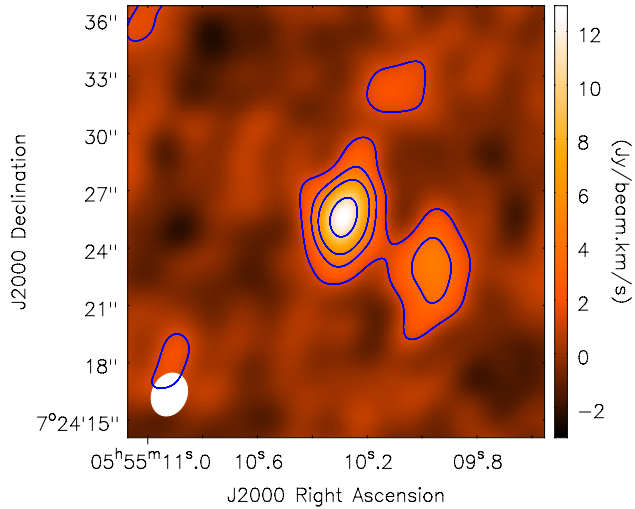}
\caption{Integrated intensity image of the D configuration channel maps that contain the discrete second source approximately 5$\arcsec$ $\rm{S-W}$ of $\alpha$ Ori. Contours for the integrated intensity are 1$\sigma$, 1.5$\sigma$, 2$\sigma$, and 3$\sigma$ (1$\sigma$ = 1.3 Jy beam${}^{-1}\>{\rm km\>s}^{-1}$). The size of the restoring beam is shown in white in the bottom left corner.}
\label{fig:fig4}
\end{figure}

An additional spatially unresolved source is detected in a number of the D configuration image cube maps (both high and low spectral resolutions) and has been previously documented by \citet{2009AIPC.1094..868H}. The component is present in only five continuous low-resolution channels between $\sim$ $-4.0\>{\rm km\>s}^{-1}$ and $+2.4\>{\rm km\>s}^{-1}$ and is located $\sim$ 5$\arcsec$ $\rm{S-W}$ of $\alpha$ Ori as shown in Figure \ref{fig:fig4}. Its peak emission lies at $\sim$ $0\>{\rm km\>s}^{-1}$ and here it approximately equals 60$\%$ of the source peak emission. The corresponding channel maps in the E configuration image cube show extended emission out to ~8$\arcsec$ in the same $\rm{S-W}$ direction. This second source does not appear in any of the C configuration channel maps probably due to the lower sensitivity resulting from the smaller HPBW (i.e., the flux is diluted). This discrete second source thus has the effect of adding extra emission to the corresponding multi-configuration image cube maps at the low velocities where it is present.

\subsection{Determination of the S1 and S2 Radii} \label{results3} 
The spatial extent of the S1 and S2 flows around Betelgeuse was not directly determined from either the CO infrared absorption spectra of \cite{1979ApJ...233L.135B} or previous CO single-dish radio observations \citep{1980ApJ...242L..25K, 1987ApJ...313..400H, 1994ApJ...424L.127H}. Our low spectral resolution multi-configuration image cube has sufficient spatial resolution and signal to noise to make direct estimates of the maximum radius of both flows. The outer S2 flow is not seen in the low absolute velocity channel maps where its spatial extent is maximum and either lies outside of the primary beam or is lost into the noise near the edge of the maps. We derive the maximum outer scale of the S2 flow by looking at the spatial scales of the S2 flow in the higher absolute velocity maps where S2 is present. If we assume that S2 is spherically symmetric with an outer radius $R_{\rm{S2}}$, and is undergoing steady expansion with velocity $V_{\rm{S2}}$, then we can estimate its radius in each velocity channel using the following relation:

\begin{equation}
r_{\rm{chan}}=R_{\rm{S2}} \rm{sin}\left[\rm{cos}^{-1}\left(\frac{\textit{v}_{chan}}{\textit{V}_{S2}}\right) \right]
\end{equation} 
where ${r_{\rm{chan}}}$ is the S2 radius in a channel at velocity $v_{\rm{chan}}$. 

\begin{figure}
\epsscale{1.2}
\includegraphics[trim=45pt 0pt 80pt 10pt, width=7.5cm, height=6.5cm]{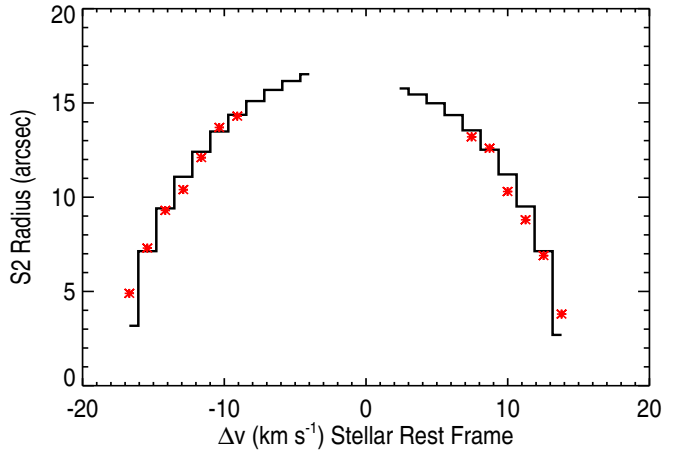}
\caption{Derived S2 radius as a function of velocity (red points) overplotted with two model outflows. The blueshifted model (left) corresponds to an outflow with a maximum radius of 17$\arcsec$ and a velocity of $16.7\>{\rm km\>s}^{-1}$ while the redshifted model (right) corresponds to an outflow with a maximum radius of 16$\arcsec$ and a velocity of $13.8\>{\rm km\>s}^{-1}$. Note: the line profile is 1.9 $1.9\>{\rm km\>s}^{-1}$ wider in the low resolution image cube ($\Delta v$ = $1.3\>{\rm km\>s}^{-1}$) than in the high resolution image cube.}
\label{fig:fig5}
\end{figure}

We use Equation (1) to estimate the maximum projected spatial extent of S2 which occurs at zero velocity. An estimate of the S2 radius per channel ($r_{\rm{chan}}$) was found by creating annuli of increasing radius around the central emission in each relevant line channel map of the multi-configuration image cube, extracting all flux within each annulus and then plotting these fluxes against distance from the star for each channel. The maximum of these resultant curves was then deemed to be the maximum radius of S2 per channel. Figure \ref{fig:fig5} shows these data over-plotted with two model outflows which were created using Equation (1). The blueshifted data points were best fitted by a model outflow of maximum radius 17$\arcsec$ and outflow velocity $17\>{\rm km\>s}^{-1}$, while the redshifted data points were best fitted by a model outflow of maximum radius 16$\arcsec$ and outflow velocity $14\>{\rm km\>s}^{-1}$. It is worth mentioning that this estimate for the spatial extent of S2 is only weakly dependent on our adopted radial velocity value for Betelgeuse and adopting a slightly different value would simply alter S2s outflow velocities. As S2 is not present in our lowest absolute velocity map we are not able to report an estimate of its width.

\begin{figure*}[hbt!]
\mbox{
          \includegraphics[scale=0.50]{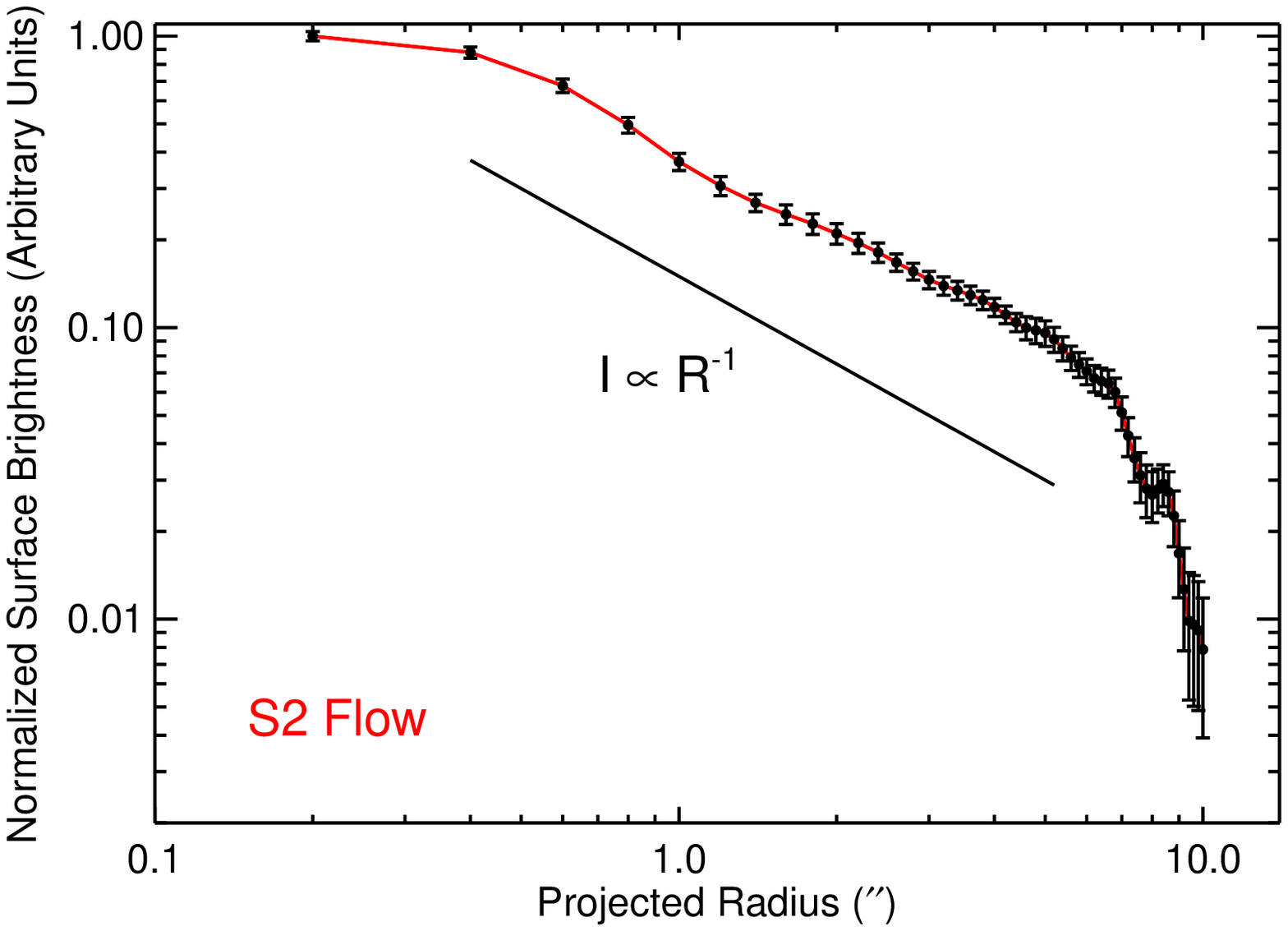}
          \includegraphics[scale=0.50]{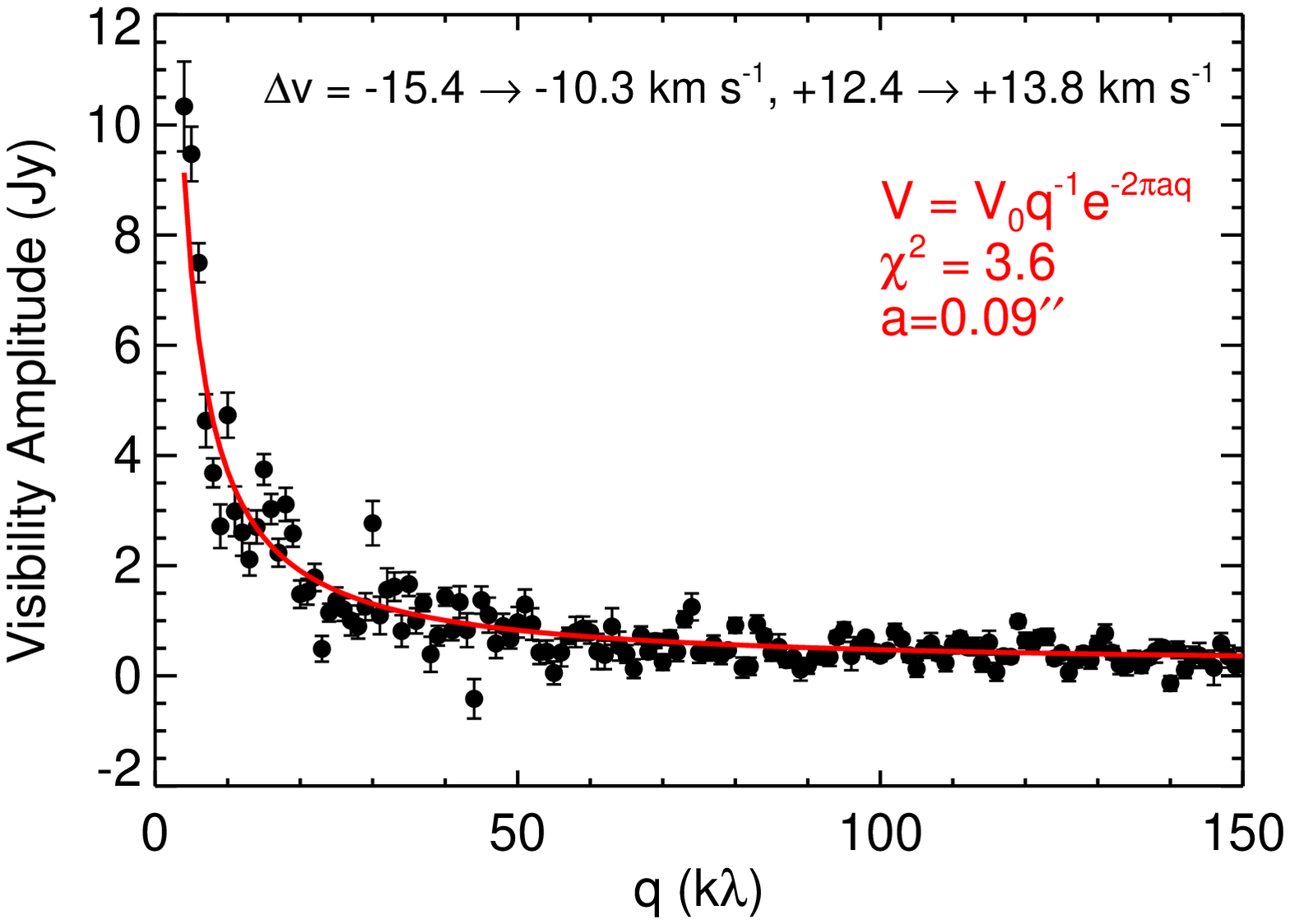}
          }
\\
\mbox{
          \includegraphics[scale=0.50]{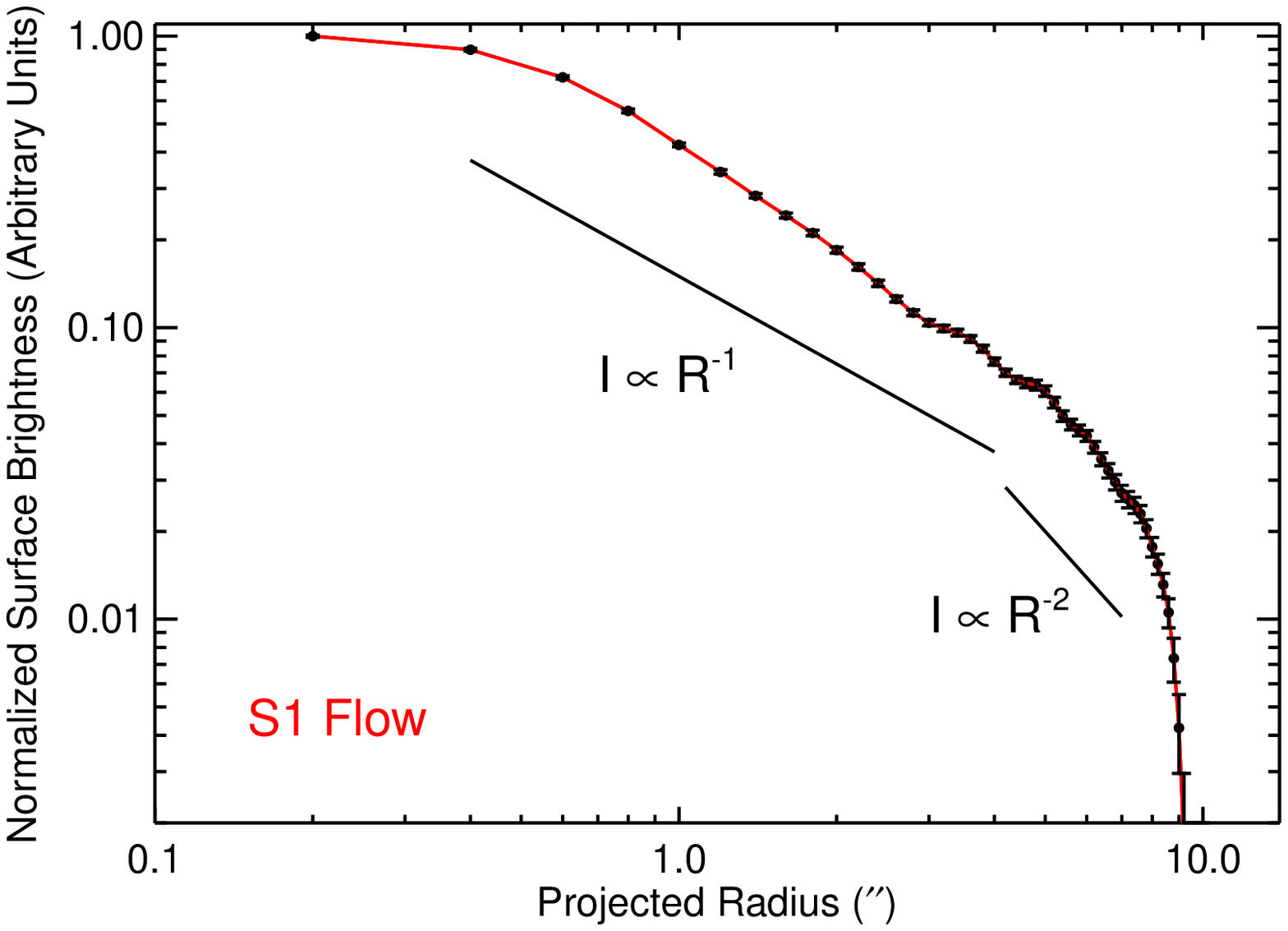}
          \includegraphics[scale=0.50]{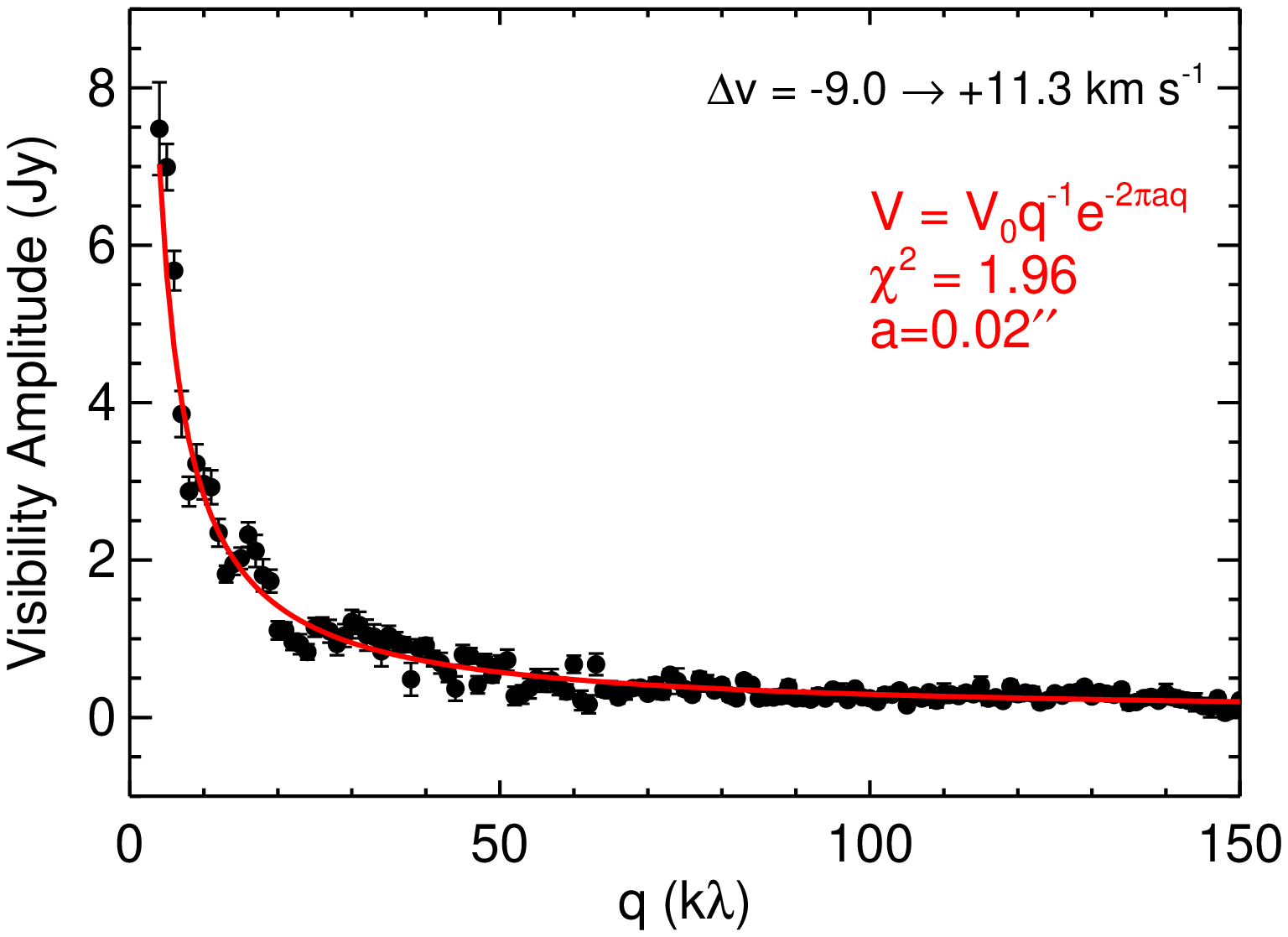}

          }
\\
\caption{Left Column: surface brightness as a function of projected radius on sky, $R$ (red line). The emission has been extracted from the low spectral resolution multi-configuration image cube and is integrated over the channels where S1 is present (bottom) and over the channels where only S2 is present (top). Intensity proportional to $R{}^{-1}$ and $R{}^{-2}$ is also shown for comparison. Right Column: the corresponding visibility amplitude as a function of $u-v$ distance ($q$) of both outflows can be modeled well by an $R{}^{-1}$ fall off in intensity. The error bars in all plots represent the standard error of the mean.}
\label{fig:fig6}
\end{figure*}

In the left column of Figure \ref{fig:fig6} we investigate the intensity distribution of CO emission as a function of projected radius, $R$, for both the S1 and S2 flows. From our discussions in Section \ref{results1} we can assume that all line emission between $-15.4 \rightarrow$ $-10.3\>{\rm km\>s}^{-1}$ and +12.4 $\rightarrow$ $+13.8\>{\rm km\>s}^{-1}$ emanates solely from the S2 flow. Using the low spectral resolution multi-configuration image cube we integrate the surface brightness over these channels and find that the intensity fall off is proportional to $R{}^{-1}$ (Figure \ref{fig:fig6}, top). To investigate the S1 flow intensity distribution around $\alpha$ Ori we integrate the surface brightness over the channels between $-9 \rightarrow$ $+10.6\>{\rm km\>s}^{-1}$. Although these channels contain emission from both S1 and S2, most of the S2 emission here will have larger projected radii and thus the majority of the inner emission should emanate from the S1 flow.  Between 0$\arcsec$.5 and 4$\arcsec$ from the star the intensity is again found to be proportional to $R{}^{-1}$ (Figure \ref{fig:fig6}, bottom). Such an intensity distribution is expected for an optically thin homogeneous constant velocity outflow with $\rho \propto 1/R^{2}$. Beyond 4$\arcsec$ the intensity fall off is more rapid and is close to an $R{}^{-2}$ distribution which may mark the initiation of the current epoch of mass-loss.

Insight can also be gained into how the intensity varies on different size scales by conducting analysis in the $u-v$ plane and plotting the visibility amplitude of $\alpha$ Ori against $u-v$ distance. The result of this is shown in the right column of Figure \ref{fig:fig6} where the same channels corresponding to the S1 and S2 flows have been used. The data are azimuthally averaged, and have been binned to produce one data point per k$\lambda$. The result for both the S1 and S2 data is a steep drop-off in visibility amplitude over a relatively short $u-v$ distance, signaling that the sources are well resolved. Both sets of visibility data agree with an intensity proportional to $(a^2 + R^2){}^{-1/2}$, where $a$ is an inner spatial limit. This is because the Hankel transform of this function is $q^{-1}e^{-2\pi aq}$ \citep{2000fta..book.....B}, where $q$ is the $u-v$ distance, and a vertically scaled version of this function is shown to match the visibility data very well in Figure \ref{fig:fig6}. As analysis in both the sky and $u-v$ plane indicate the intensity of both flows is proportional to $R{}^{-1}$ we conclude that when azimuthally averaged, both outflows are  consistent with an optically thin and quasi-steady flow which is in agreement with \cite{2009AJ....137.3558S} (i.e., S1) and \cite{2002A&A...386.1009P} (i.e., S2). 

An exact determination of the maximum spatial extent of the S1 flow is more difficult as we do not see the classical shell formation signature for it as we sample across velocities, like we do for S2. Instead its spatial extent varies over the channel maps with evidence of discrete clumps being present in many of these maps. At 20\% of maximum emission in the integrated intensity S1 map (i.e., composed of all channels between $-9 \rightarrow$ $+10.6\>{\rm km\>s}^{-1}$) the S1 flow extends out to a mean distance of $\sim$ 4$\arcsec$ and is even more extended in the $\rm{S-W}$ direction due to the presence of the second emission feature in the compact configuration data sets. The HPBW of 0$\arcsec$.9 is not sufficient to determine whether the S1 flow is discrete or an extension of the current wind phase seen in ultraviolet spectra, e.g., \cite{1997ApJ...479..970C}, and centimeter-radio continuum interferometry \citep{1998Natur.392..575L, harper_2001}.

\subsection{Continuum Flux Densities} \label{results4} 

In Table 2 we show the derived continuum flux densities for each of the three configuration image cubes and also the multi-configuration image cube. The high spectral resolution ($\Delta v$ = $0.65\>{\rm km\>s}^{-1}$) image cubes were just wide enough to image the CO line but were too narrow to make accurate estimates of the continuum flux density. Therefore, all continuum flux density estimates are derived from the lower spectral resolution ($\Delta v$ = $1.3\>{\rm km\>s}^{-1}$) image cubes from which we were able to take accurate measurements at both sides of the line. We fitted elliptical Gaussians to $\sim$ 20 continuum channels using CASA's \textit{imfit} routine allowing the flux and corresponding uncertainties to be calculated. The source was unresolved in most of these continuum channels. 

Betelgeuse is known to show brightness variations at many wavelengths. \cite{1984PASP...96..366G} reports a decrease of half a magnitude in visual brightness over a period of six years. \cite{1987LNP...291..337B} found stochastic 30\%-40\% variations in flux density at 6\,cm over timescales as short as 10 days to as long as 8 months (i.e., the observational period). A more comprehensive study was carried out by \cite{1992ASPC...26..455D} who observed Betelgeuse with the VLA at centimeter-wavelengths from 1986 to 1990 and found stochastic variability of 22\%, 15\%, and 21\% at 6\,cm, 3.6\,cm, and 2\,cm, respectively at a variety of different timescales down to less than 1 month. The millimeter-continuum emission that we measure arises mainly from electron$-$ion and electron$-$atom bremsstrahlung and possibly dust emission, so it is not unreasonable to also expect variability at millimeter-wavelengths too. The D configuration data were acquired under adverse weather conditions and these data have the highest noise levels out of the three configurations. Its continuum emission measurement is approximately 50\% greater than the C and E configuration continuum measurements which were also acquired approximately two years after the D configuration data.  We believe that the continuum emission derived from the multi-configuration image cube is a reasonable estimation of the mean millimeter-continuum flux density over the two year period and is in reasonably good agreement with the 250\,GHz flux density of \cite{1994A&A...281..161A} who report a mean value of 351$\pm$25\,mJy for 1986 $\rightarrow$ 1989.

\begin{deluxetable}{cccc}
\tabletypesize{\scriptsize}
\tablecaption{CARMA Continuum Fluxes at 230 GHz}
\tablewidth{0pt}
\tablehead{
\colhead{Configuration} 		&
\colhead{Restoring Beam }		&
\colhead{Flux}			&
\colhead{Uncertainty}		\\
\colhead{}			&
\colhead{($\arcsec \times \arcsec$)}   	&
\colhead{(mJy)}			&
\colhead{(mJy)} 
}
\startdata
C & 0.96 $\times$ 0.76 & 234 & 18\\
D & 2.33 $\times$ 1.87 & 389 & 72\\
E & 4.93 $\times$ 3.84 & 278 & 40 \\
Multi-configuration & 1.05 $\times$ 0.84 & 289 & 21
\enddata
\label{tab:tab2}
\end{deluxetable}
\section{DISCUSSION}
\subsection{Previous CO Observations}
\cite{1979ApJ...233L.135B} were the first to detect circumstellar absorption lines in CO by looking at the fundamental ro-vibration lines at 4.6$\,\mu$m. These infrared observations revealed two separate outflows around $\alpha$ Ori; an excited (\rm{$T_{\rm{exc}}$}\,= 200\,K) S1 flow with an expansion velocity of $9\>{\rm km\>s}^{-1}$ and a less excited (\rm{$T_{\rm{exc}}$} = 70 K) S2 flow moving with a faster expansion velocity of $16\>{\rm km\>s}^{-1}$. \cite{1980ApJ...242L..25K} were the first to detect emission in the CO($J=$ $2-1$) line at 1.3 mm using the 10 m millimeter-wave telescope at OVRO but only detected one component expanding at $15\>{\rm km\>s}^{-1}$. By analyzing the shape of the line profile, they concluded that the S2 radius of 55$\arcsec$ derived by \cite{1979ApJ...233L.135B} was too large and that it lies at a radius of $R \leq$ 10$\arcsec$. Since the detection by Knapp, a number of observations at 1.3 mm have been carried out with various beam sizes and all spectra look remarkably similar; that is the profile has a steep extreme blue shifted emission component with the remainder of the profile looking more flat-topped and containing a number of less dominant spikes. \cite{1987ApJ...313..400H} used their single-dish observations (HPBW $\sim$ 32$\arcsec$) of the CO($J=$ $2-1$) line along with excitation and self-shielding models of CO to conclude that the S1 flow makes little contribution to the final emission line. They also identify the extreme blue wing of the line with the S2 flow and predict that it may extend out to a radius of $\sim$ 16$\arcsec$. Later however, \cite{1994ApJ...424L.127H} compared their detected 609 $\mu$m ${}^3P{}_1\rightarrow{}^3P{}_0$ fine structure line of C I with CO data obtained with the IRAM 30\,m telescope (HPBWs $\sim$ 12$\arcsec$) and find that the expansion velocities in both lines are essentially the same. They conclude that the radial extent of C I is $\lesssim$ 7$\arcsec$ and both the CO and C I are formed in the inner envelope and roughly extend over the same area.  

The shape of our multi-configuration line profile for extraction areas of radii 6$\arcsec$ or greater are in good agreement with previous high signal to noise single-dish CO($J=$ $2-1$) spectra \cite[e.g.][Figure 1]{1994ApJ...424L.127H} although the emission spikes in our line profiles are more dominant. Our total line width of $28.6\>{\rm km\>s}^{-1}$ is in good agreement with \cite{1987ApJ...313..400H} and \cite{1994ApJ...424L.127H} who report line widths of $28.6\>{\rm km\>s}^{-1}$ and $30\>{\rm km\>s}^{-1}$ respectively. The extreme blue wing in both of these spectra is the dominant emission feature of the line and this is also true in our multi-configuration spectra at extraction areas $\gtrsim$ 6$\arcsec$. Using the IRAM 30\,m telescope, which has an HPBW of only 12$\arcsec$ at 230 GHz, \cite{1994ApJ...424L.127H} produce a similar line profile shape to that presented in \cite{1987ApJ...313..400H} who have a larger HPBW of $\sim$30$\arcsec$. From this, one would expect that the majority of the blue wing emission is compact. Our multi-configuration line profiles suggest otherwise however, and show a continuous increase in the blue wing emission as we take larger extraction regions out to 10$\arcsec$. The multi-configuration maps also show a faint ring structure forming at $\sim$ $-11.6\>{\rm km\>s}^{-1}$ and expanding further out in lower absolute velocity channel maps. This ring emission is fainter than the higher velocity compact emission so we see a drop in flux density in our spectra at the point where these rings form. Therefore, the steepness of the extreme blue wing in our multi-configuration spectrum means that there is merely more CO emitting at higher velocities than at lower, which is indicative of a sheet like structure moving towards the observer.

The line profiles of higher CO rotational transitions for Betelgeuse have been published in \cite{2003A&A...407..609K} and \cite{2010A&A...523A..18D}. \cite{2010A&A...523A..18D} present high spectral resolution (0.3125 MHz) line profiles for the CO($J=$ $2-1$), ($J=$ $3-2$), and ($J=$ $4-3$) transitions that were obtained with the James Clerk Maxwell Telescope (JCMT). For the CO($J=$ $2-1$) transition the JCMT has an HPBW of $\sim$ 20$\arcsec$ and the profile appears similar to our multi-configuration profile over the same flux density extraction area (i.e., Figure \ref{fig:fig2}), with the extreme blue wing component being the dominant feature in both. This feature, which is emission from the S2 flow, becomes a less dominant component of the line profile at the higher CO($J=$ $3-2$) and CO($J=$ 4-3) transitions where the JCMT has an HPBW of $\sim$ 13$\arcsec$ and 8$\arcsec$, respectively, and does not capture all of the S2 emission which is shown in Figure \ref{fig:fig3} to be extended at these velocities. Also, the higher rotational states ($J\approx 10$) will be populated more by the higher excitation temperature ($\sim$ 200 K) S1 flow so these line profiles become dominated by emission from the slower S1 flow. This is confirmed by our narrow SOFIA-GREAT CO($J=$ $12-11$) line profile which is presented in the appendix of this paper, and also by a visual inspection of Herschel-HIFI archival line profiles of the CO($J=$ $6-5$), ($J=$ $10-9$), and ($J=$ $16-15$) transitions.

The CO 4.6 $\mu$m ro-vibration lines have been observed with the Phoenix spectrograph \citep{1998SPIE.3354..810H} by \cite{1999A&A...347L..35R} and \cite{2009AJ....137.3558S} on the 2.1 m telescope at Kitt Peak and on the 8 m Gemini South telescope respectively. By assuming a Boltzmann population distribution for the ground rotational levels of CO, \cite{1999A&A...347L..35R}  derived a mean excitation temperature of 38${}^{+6}_{-5}$ K along the line of sight at a projected distance of 4$\arcsec$ north of Betelgeuse. Our CARMA data suggests that the S1 flow extends out to approximately this distance but Ryde et al.'s temperature is not in agreement with either of the line-of-sight S1 or S2 excitation temperatures of 200${}^{+50}_{-10}$\,K and 70$\pm 10$\,K derived by \cite{1979ApJ...233L.135B}. This discrepancy may indicate that the excitation is quite non-uniform.
\cite{2009AJ....137.3558S} did not derive an excitation temperatures but used their 4.6 $\mu$m spectra to reveal extended resonantly scattered CO emission out to $\sim$ 2$\arcsec$, a factor of two smaller than our S1 radius. They observe emission over a velocity range of 30 km s${}^{-1}$ but two distinct flows are not detected. Mild ($\sim$ 20\%) density inhomogeneities are reported but overall, their observations are consistent with an optically thin and steady wind which is consistent with our findings.

\subsection{K\,I\,7699\,\AA \ Spectra}

The S2 flow was first identified in high resolution K\,I and Na\,I absorption spectra by \cite{1975ApJ...199..427G} and subsequently re-observed multiple times over the next couple of years \citep{1979QJRAS..20..361G}. It is interesting to compare these and Bernat et al.'s (1979) CO line-of-sight absorption velocities with those from the CARMA emission spectra obtained at similar spectral resolutions and also to measure, the perhaps co-spatial line broadening of the K\,I S2 absorption feature.

We have obtained K\,I 7698.98 \AA \ spectra using the cross-dispersed echelle spectrometers on the Harlan J. Smith 107 inch (2.7m) reflector at McDonald Observatory. With 2 pixels per resolution element an $R=\lambda/\Delta\lambda=200,000$ and a $R=500,000$ spectrum were obtained on 2007 March 25 and April 13, respectively. The spectra were wavelength calibrated with ThAr lamp lines and the lower resolution spectrum was checked by fitting six symmetric terrestrial O${}_2$ lines in the same order using wavelengths from \cite{1948ApJ...108..167B}. The O${}_2$ lines confirmed that the $R=200,000$ calibration was good to better than $0.1\>{\rm km\>s}^{-1}$. Upon cross-correlating the low- and high-resolution spectrum the high-resolution spectrum appeared redshifted by $0.60\>{\rm km\>s}^{-1}$, i.e., one resolution element, for which we do not have an explanation except to note that a similar offset has been reported by \cite{1994ApJ...436..152W}. We use the cross-correlation to define the wavelength calibration of the $R=500,000$ spectrum and we adopt a systematic error of $\sigma^{\rm{sys}}=0.2\>{\rm km\>s}^{-1}$.

\begin{figure}
\epsscale{1.0}
\includegraphics[trim=40pt 0pt 50pt 0pt, width=8.0cm, height=7.5cm]{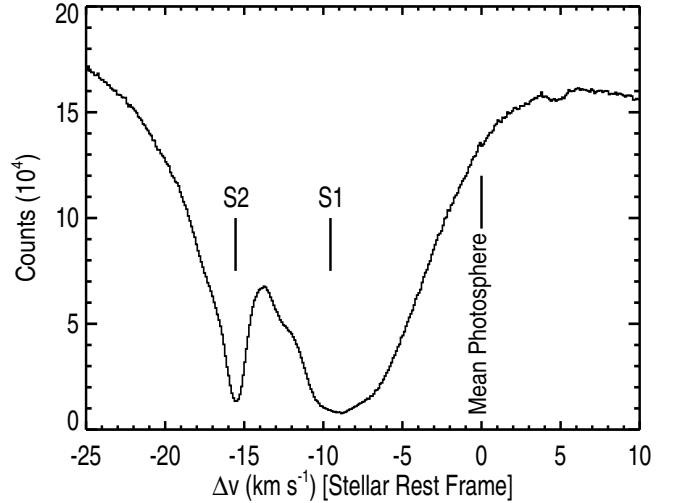}
\caption{K\,I 7698.98 \AA \  profile ($R$=500,000) for Betelgeuse obtained on 2007 April 13. The S2 flow outflow velocity is found to be 15.6 km s${}^{-1}$ (using a $V_{\rm{rad}}$ = +20.7 km s${}^{-1}$), slightly lower than \citeauthor{2002A&A...386.1009P}'s \citeyearpar{2002A&A...386.1009P} value of 18$\pm$2 km s${}^{-1}$.}
\label{fig:fig7}
\end{figure}

The high-resolution spectrum is shown in Figure \ref{fig:fig7} in the adopted stellar center-of-mass rest frame ($V_{\rm{rad}}=+20.7\>{\rm km\>s}^{-1}$). The S2 feature is deep, well separated from the S1 feature, and very well represented by an absorption model including the hyperfine splitting from the abundant $^{39}K$ isotope. We adopt the K\,I $7698.9645$ $\rm{\AA}$ line parameters compiled in \cite{2003ApJS..149..205M}\footnote{Note that this wavelength was recently revised and is $0.44\>{\rm km\>s}^{-1}$ less that that adopted in the Goldberg studies.} and find a heliocentric S2 absorption velocity of $+5.15\pm 0.01(1\sigma^{\rm{ran}})\>{\rm km\>s}^{-1}$ with the uncertainty dominated the absolute wavelength scales. However, the most probable line-of-sight turbulent velocity is much more accurately determined with $0.62\pm 0.02(1\sigma^{\rm{ran}}) \>{\rm km\>s}^{-1}$. From their 4.6 $\mu$m line profiles Bernat et al. (1979) found a very similar $1\>{\rm km\>s}^{-1}$, these being the sum of thermal and non-thermal motions. These values place tight constraints on the line-of-sight 
acceleration associated with the S2 flow. 

The K\,I spectrum also reveals a slight inflection in the observed line profile at $+3.6\>{\rm km\>s}^{-1}$ (heliocentric) which may represent structure in the underlying photospheric profile or additional absorption in which case it has $\sim 0.1$ the column density of S2 ($N_{\rm{KI-S2}}\simeq 1.2\times 10^{11}\>{\rm cm}^{-2}$).  The S2 absorption velocity minima can be compared to those obtained by \citet[Figure 7]{1979QJRAS..20..361G} who measured values between 1975 and 1978 of $4.2\pm 0.2$ and $5.0\pm 0.2 \>{\rm km\>s}^{-1}$ and these differences may result from changes caused by radial velocity changes in the underlying photospheric spectrum. Bernat et al.'s 1979 CO IR absorption observations reveal S2 heliocentric velocities of $+4.94\pm 0.30$ $\>{\rm km\>s}^{-1}$ (1979 March 6) and $+4.60\pm 0.04 \>{\rm km\>s}^{-1}$ (1979 April 14) with turbulent velocities of 4 and $1\>{\rm km\>s}^{-1}$ for the S1 and S2 features, respectively.

In terms of the center-of-mass radial velocity of the star our K\,I feature implies an outflow velocity of $+15.6\>{\rm km\>s}^{-1}$. The blue edge of our CARMA multi-configuration CO profile is estimated to be $+15.4\>{\rm km\>s}^{-1}$  which suggests a dynamical association with the CO S2 flow and very close agreement with \citeauthor{1979ApJ...233L.135B}'s \citeyearpar{1979ApJ...233L.135B} CO absorption velocities listed above. \cite{2002A&A...386.1009P} have also estimated the radius and velocity of the suspected K\,I S2 flow using $R=110,000$ resolution long slit spectra. They found a geometrically thin shell (1$\arcsec$) with velocity of $V_{\rm{S2}}=18\pm 2\>{\rm km\>s}^{-1}$ with a radius of 55$\arcsec$ which is much larger than the field of view of the CARMA spectra. Their long slit spectra show several smaller partial shells but it is not simple to directly associate the CO emission feature with one or more of these shells especially given the uncertainty in the ionization balances of CO and K\,I. It is possible that the 55$\arcsec$ shell is associated with the inflexion caused by additional absorption (and low column density) at a velocity slightly higher then S2 observed in our K\,I profile.

\section{CONCLUSIONS}
The two distinct velocity components seen by \cite{1979ApJ...233L.135B} in CO absorption against the stellar spectrum at 4.6 $\mu$m have both been detected at 230 GHz for the first time. The first velocity component known as S1 has an expansion velocity of $9\>{\rm km\>s}^{-1}$ \citep{1979ApJ...233L.135B} and is detected in our high spectral resolution C configuration profile with the same blueshifted velocity (i.e., $-9.0\>{\rm km\>s}^{-1}$) and with a larger redshifted outflow velocity of $+10.6\>{\rm km\>s}^{-1}$. The extended CARMA C configuration has a resolving out scale of $\sim$ 6$\arcsec$ and thus spatially filters almost all of the S2 emission leaving us with an approximate spectrum for the S1 flow. An extreme blue wing of the CO spectrum appears in the D and E configuration spectra which we associate with the S2 flow. The high spectral resolution multi-configuration spectrum is used to determine S2 outflow velocities of $-15.4\>{\rm km\>s}^{-1}$ and $+13.2\>{\rm km\>s}^{-1}$ which is in good agreement with our K\,I\,7699\,\AA \ line-of-sight S2 velocity and that reported by \cite{1979ApJ...233L.135B}. 

The low spectral resolution multi-configuration maps provide the first direct measurements on the spatial extent of the S2 flow, which we derive to have a radius of 17$\arcsec$; a value that is higher than most previous estimates. We do not see a well defined outer edge for the S1 flow but believe that it may extend out to a radius of $\sim$ 4$\arcsec$. Previous single-dish observations of the CO line with small HPBWs do not show the classical resolved signature of high emission at large absolute velocities and low emission at low absolute velocities for two main reasons. First, the S1 flow is still unresolved in these single-dish observations and thus contributes emission and at the lower absolute velocities. As well as this, the multi-configuration CARMA maps show that the S2 emission is brighter in the higher absolute velocity maps than at lower absolute velocities and so when the emission from the fainter rings is neglected (i.e., when observed with a small HPBW), the overall line profile does not change significantly.

Assuming a mean outflow velocity of $14.3\>{\rm km\>s}^{-1}$ and $9.8\>{\rm km\>s}^{-1}$ for the S2 and S1 flows respectively then their ages are $\sim$ 1100 yr and $\sim$ 380 yr. Since \cite{2002A&A...386.1009P} have detected K\,I out to 55$\arcsec$ at a similar velocity to the CO S2 flow, then, assuming the CO and K\,I are coupled, there appears to be little or no further acceleration in Betelgeuse's outflow once the S2 flow begins (which is somewhere greater than 4$\arcsec$). The composition and dynamics of the interface between S1 and S2 remains unknown and future instruments such as the Atacama Large Millimeter/submillimeter Array (ALMA) will provide a greater understanding of this region. Higher spatial resolution, increased sensitivity, and excellent $u-v$ coverage are needed to determine if the inner S1 flow is discrete or just an extension of the current wind phase. Our SOFIA-GREAT spectrum shows that the higher excitation gas traces the slower S1 component, and therefore the high frequency bands of ALMA will preferentially trace the S1 flow. Solutions to these remaining puzzles will broaden our knowledge of the evolutionary aspect of Betelgeuse's outflow and shed light into the driving mechanisms of M supergiant winds.

\acknowledgments

Support for CARMA construction was derived from the states of California, Illinois, and
Maryland, the James S. McDonnell Foundation, the Gordon and Betty Moore Foundation, the
Kenneth T. and Eileen L. Norris Foundation, the University of Chicago, the Associates of the
California Institute of Technology, and the National Science Foundation. Ongoing CARMA
development and operations are supported by the National Science Foundation under a
cooperative agreement, and by the CARMA partner universities. This work is based (in part) on observations made with the NASA/DLR Stratospheric Observatory for Infrared Astronomy. SOFIA Science Mission Operations are conducted jointly by the Universities Space Research Association, Inc., under NASA contract NAS2-97001, and the Deutsches SOFIA Institut under DLR contract 50 OK 0901. GREAT is a development by the MPI f\"ur Radioastronomie and the KOSMA/ Universit\"at zu K\"oln, in cooperation with the MPI f\"ur Sonnensystemforschung and the DLR Institut f\"ur Planetenforschung. We thank Sarah Kennelly for providing us with archival Herschel spectra. This publication has emanated from research conducted with the financial support of Science Foundation Ireland under Grant Number SFI11/RFP.1/AST/3064, and a grant from Trinity College Dublin.

{\it Facilities:} \facility{CARMA}, \facility{Smith} and \facility{SOFIA}.

\bibliography{references}

\appendix
\section{SOFIA-GREAT OBSERVATION}
As part of our larger multi-wavelength study of the CO surrounding Betelgeuse we observed the star (PI: Harper; ID 81\_0005\_1) with the German Receiver for Astronomy at Terahertz Frequencies \cite[GREAT;][]{2000SPIE.4014...23G} instrument on NASA and DLR's Stratospheric Observatory for Infrared Astronomy \cite[SOFIA;][]{2009ASPC..417..101B} 2.5m airborne observatory. The $\rm{^{12}C{}^{16}O}$(\textit{J} = $12-11$, 1.38~THz, $216.9\,\rm \mu m$) line was observed to examine the dynamics of the higher excitation S1 component. The observations were made during Flight 86 on 2011 November 10 at 13,100 m ($\sim$ 43,000 ft) when the star had an elevation of 45$\degr$. The HPBW was $\sim 19\arcsec$ and the effective on source exposure time was 12 minutes. Due to technical difficulties during the flight the observations were obtained using a non-standard asymmetric chop sequence with a throw of $60\arcsec$.

\begin{figure*}[hbt!]
\centering
\includegraphics[trim=60pt 0pt 0pt 0pt, scale=0.5]{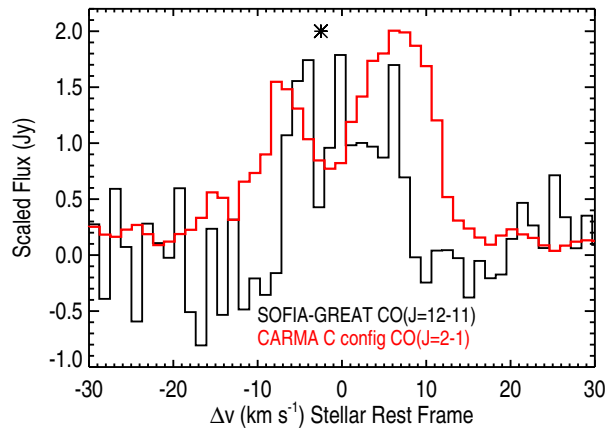}
\caption{Scaled SOFIA-GREAT CO(\textit{J} = $12-11$) emission line profile binned to $1.3\>{\rm km\>s}^{-1}$ along with the unscaled CARMA C configuration CO(\textit{J} = $2-1$) profile. The star symbol at $-2.5\>{\rm km\>s}^{-1}$ marks the location of a weak terrestrial atmospheric emission feature which may contribute to the drop in flux at this velocity in the SOFIA-GREAT line profile.}
\label{fig:fig8}
\end{figure*}

The scaled SOFIA-GREAT emission line profile binned to $1.3\>{\rm km\>s}^{-1}$ is shown in Figure \ref{fig:fig8} along with the CARMA C configuration CO(\textit{J} = $2-1$) profile which spatially filters the S2 contribution. This figure shows that the $\textit{J} = 12-11$ profile reflects the slower moving S1 flow with a width of $\sim \pm 7.5\>{\rm km\>s}^{-1}$ approximately centered on the stellar rest frame. This suggests that the $\textit{J} = 12-11$ emitting plasma is not associated with the faster S2 component and is likely associated with the higher excitation S1 plasma. Owing to uncertainties in the pointing accuracy during our GREAT observation we defer a discussion of the fluxes to a later time. 

\end{document}